# A Tale of Two Mobile Generations: 5G-Advanced and 6G in 3GPP Release 20

Xingqin Lin

NVIDIA

Email: xingqinl@nvidia.com

*Abstract*—As the telecommunications industry stands at the crossroads between the fifth generation (5G) and sixth generation (6G) of mobile communications, the 3rd generation partnership project (3GPP) Release 20 emerges as a pivotal point of transition. By striking a balance between enhancing 5G-Advanced capabilities and setting the stage for 6G, Release 20 provides the crucial foundation upon which future mobile communication standards and deployments will be built. This article examines these dual objectives, outlining the key enhancements, the motivations behind them, and their implications for the future of mobile communications.

## I. INTRODUCTION

The evolution of mobile communication networks has been marked by periodic leaps in performance, capabilities, and the breadth of supported use cases. The fifth generation (5G) new radio (NR), standardized by the 3rd generation partnership project (3GPP) starting with Release 15, significantly broadened the scope of mobile connectivity [1]. 5G can operate in either non-standalone (NSA) mode, leveraging long-term evolution (LTE) for initial access and mobility, or standalone (SA) mode, which is independent of LTE. It introduced a flexible framework capable of supporting enhanced mobile broadband (eMBB), ultra-reliable low-latency communications (URLLC), and massive machine-type communications (mMTC).

Over subsequent Releases 16 and 17, 5G evolution adopted new paradigms, including non-terrestrial networks (NTNs) and integrated access and backhaul (IAB) [2]. The evolution of 5G-Advanced began with Release 18 [3]. As 3GPP progresses through Releases 18 and 19, 5G-Advanced takes shape, emphasizing enhancements in energy efficiency, addressing emerging requirements from applications such as extended reality (XR) services and ambient internet of things (IoT), and exploring artificial intelligence (AI)/machine learning (ML) based optimizations [4].

Release 20 stands at a unique juncture, serving as both the last major release of 5G-Advanced and the first major release of 6G studies [5]. On one hand, Release 20 focuses on a more selective set of essential enhancements for 5G-Advanced to address identified deployment needs, positioning 5G-Advanced to achieve its commercial potential. On the other hand, Release 20 investigates 6G use cases, scenarios, requirements, architecture, and enabling technologies, setting the technical cornerstones for 6G specification development in future

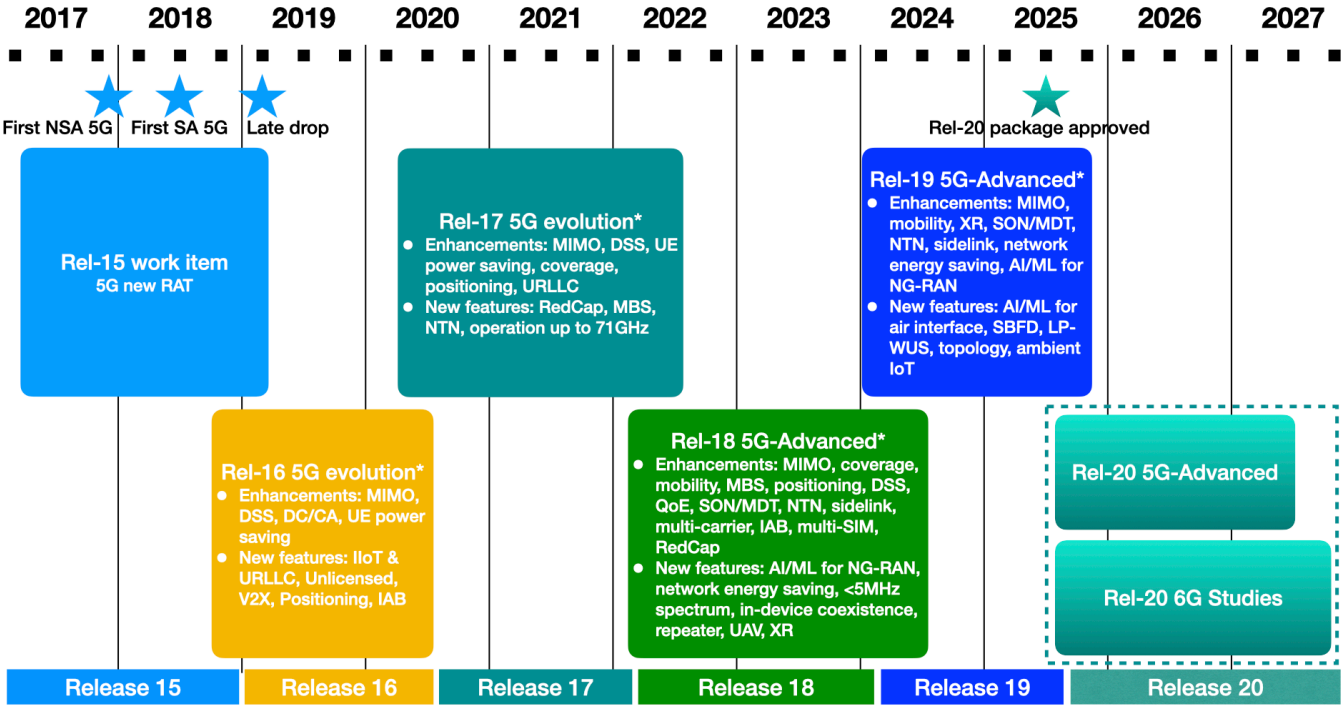

Figure 1: 3GPP's evolution roadmap from 5G to 5G-Advanced and 6G (indicative).

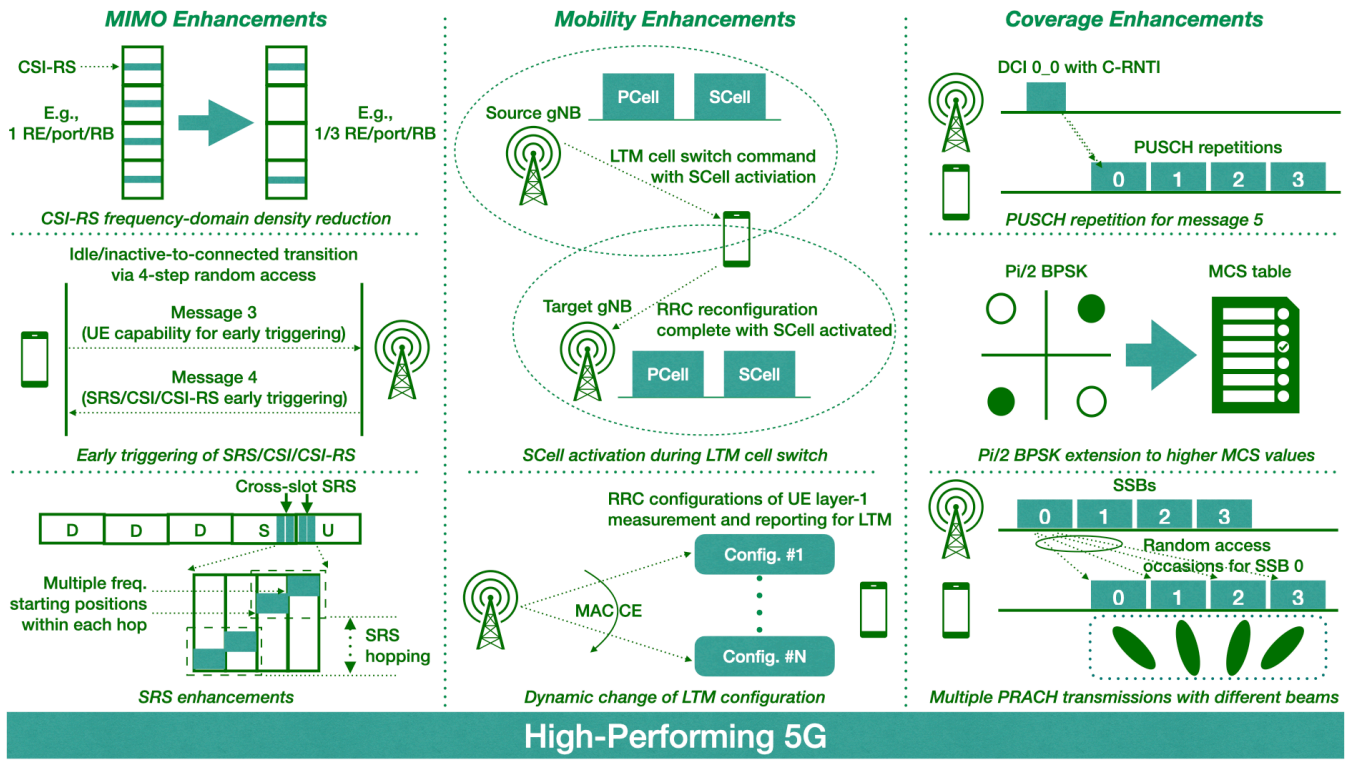

Figure 2: An illustration of the key 3GPP Release-20 features related to high-performing 5G.

Release 21. Figure 1 outlines the 3GPP's roadmap for the evolution of 5G from its inception to 5G-Advanced and 6G.

While 3GPP plays a central role in defining 6G technologies, the international telecommunication union radiocommunication sector (ITU-R) is responsible for establishing the criteria that a new radio technology must meet to qualify as an international mobile telecommunications-2030 (IMT-2030) system (i.e., 6G) [6]. ITU-R has laid out six usage scenarios for IMT-2030, including 1) immersive communication, 2) hyper-reliable and low-latency communication, 3) massive communication, 4) ubiquitous connectivity, 5) AI and communication, and 6) integrated sensing and communication (ISAC). While the first three usage scenarios are extensions from 5G eMBB, URLLC, and mMTC usage scenarios, respectively, the last three are new usage scenarios envisioned for 6G. ITU-R also governs the global spectrum alignment process, building consensus on the spectrum allocations necessary for cellular technologies. Recognizing the importance of alignment, 3GPP has been actively engaging with ITU-R and harmonizing its 6G efforts with the IMT-2030 framework.

In this article, we explore the dual role of 3GPP Release 20 as both a capstone in the evolution of 5G-Advanced and a stepping stone toward 6G. On one hand, we examine the key directions set forth in Release 20 for 5G-Advanced and describe how the selected enhancements not only meet current industry demands but also anticipate future opportunities. On the other hand, we discuss the 6G study items in Release 20, with a particular focus on radio access network (RAN) scenarios, requirements, and technologies. In doing so, this article provides a timely overview of the technological continuum spanning from 5G-Advanced to the nascent outlines of 6G, offering insights into how 3GPP Release 20 ensures the continued relevance of 5G-Advanced while making steady progress toward 6G.

## II. HIGH-PERFORMING 5G

3GPP further improves 5G performance in Release 20 by enhancing multiple-input multiple-output (MIMO) technique, mobility, and coverage, as illustrated in Figure 2.

### A. MIMO Enhancements

The widespread adoption of MIMO in mobile communication networks gained momentum with the advent of LTE. Over time, MIMO has evolved into what is now known as massive MIMO, a cornerstone of 5G technology. 3GPP has continuously enhanced MIMO support across multiple releases [7]. In Release 20, MIMO enters its sixth phase of evolution for 5G, introducing several enhancements to optimize its performance and meet emerging industry needs.

One key enhancement involves the acquisition of downlink channel state information (CSI). 3GPP Release 19 introduced support for up to 128 antenna ports in NR, enabling more advanced massive MIMO configurations. However, the accompanying linear growth in CSI reference signal (CSI-RS) overhead under the legacy CSI-RS design limits performance gains and constrains the effective utilization of larger-scale antenna arrays. Accordingly, Release 20 targets reducing the CSI-RS frequency-domain density to lower values for 48, 64, and 128 CSI-RS ports, striking a balanced trade-off between CSI measurement accuracy and pilot overhead.

Release 20 also targets early CSI acquisition to enhance downlink throughput performance. In the current NR operation, the transition of a user equipment (UE), e.g., from idle to connected mode, occurs without up-to-date channel measurements. As a result, a 5G node B (gNB) must initially fall back to an open-loop transmission configuration with a conservative modulation and coding scheme (MCS) and without closed-loop precoding, resulting in a throughput drop following the state transition. To overcome this, Release 20 aims to introduce the early triggering of the sounding reference signal (SRS), CSI reporting, and CSI-RS measurements for transitions from idle/inactive to connected mode in the random access procedure, as well as for secondary cell (SCell) activation or exit from SCell dormancy in connected mode. The introduction of early triggering will enable the gNB and UE to establish accurate channel knowledge, apply closed-loop precoding, select the proper MCS, and maintain fine time-frequency tracking as soon as possible after the transition.

Uplink MIMO performance improvement via SRS enhancements is also a priority. In time-division duplex systems, cell-edge performance is often limited by the capacity and coverage of uplink SRS. Existing SRS transmissions are often confined to an 'S' slot (i.e., a slot configured with downlink symbols, uplink symbols, and guard symbols in between), typically offering only two or four uplink symbols for the SRS transmissions. Release 20 addresses this limitation by enabling SRS transmissions to span two adjacent slots (i.e., one 'S' slot and one neighboring uplink slot), thereby allocating a greater number of contiguous symbols for sounding and improving coverage at the cell edge. Besides, Release 20 introduces multiple frequency-domain starting positions for SRS repetition symbols within a partial-frequency sounding hop. This enables channel interpolation to be used, thereby achieving higher estimation accuracy and diversity gain.

*B. Mobility Enhancements*

Mobility support is a fundamental feature for high-performing 5G networks. Continuous advancements in this area have enhanced network efficiency and provided a seamless user experience. However, mobility management is a complex task due to the challenges posed by large numbers of UEs, varying speeds and movement patterns, and diverse service requirements, among others. Traditional mobility management methods are based on layer-3 measurements and radio resource control (RRC) signaling, which can lead to signaling overhead and prolonged connection interruptions.

To address these challenges, 3GPP introduced layer-1/layer-2 triggered mobility (LTM) in Release 18. The LTM process consists of three main steps: preparation, pre-synchronization, and execution [8]. In the preparation phase, the network configures the UE with LTM candidate cells and receives layer-1 measurement reports from the UE. In the pre-synchronization phase, the UE establishes downlink and uplink synchronization with the target cell while maintaining its connection to the source cell. For the downlink pre-synchronization, the network sends a medium access control (MAC) control element (CE) to the UE to activate the candidate cell's transmission configuration indication (TCI) state. For the uplink pre-synchronization, the network sends a physical downlink control channel (PDCCH) order to trigger random access. Finally, in the execution phase, the network transmits another MAC CE to trigger the UE to switch to the target cell and perform the remaining reconfiguration. Thanks to the downlink and uplink pre-synchronizations, the interruption time of LTM is shorter than that of layer-3-based handover.

In previous Releases 18 and 19, LTM already supported carrier aggregation (CA) scenarios, allowing SCells to be part of the target cell group configuration during an LTM cell-switch operation. However, these SCells are configured in a deactivated state after a cell switch. As a result, the network must subsequently activate these SCells in the target cell group after completing the cell-switch procedure. In Release 20, 3GPP specifies mechanisms that enable the network to directly trigger the activation of SCells through the LTM cell-switch procedure, thereby improving efficiency and reducing latency. In addition, 3GPP defines the necessary configuration and related procedures for using MAC CE to dynamically select among RRC configurations provided for UE layer-1 measurement and reporting under LTM scenarios.

*C. Coverage Enhancements*

Network coverage has a significant impact on service quality and operator costs. Uplink coverage tends to be a major limitation, particularly with increased usage such as video uploads. In Releases 17 and 18, 3GPP introduced coverage enhancements targeting various uplink channels. However, the current specifications do not allow physical uplink shared channel (PUSCH) repetition transmission for the RRC setup complete message (also known as message 5), which leads to coverage issues during initial access. To resolve this limitation, 3GPP Release 20 enables repetition transmission for PUSCH scheduled by downlink control information (DCI) format 0_0 scrambled with cell radio network temporary identifier (C-RNTI).

The pi/2-binary phase shift keying (BPSK) modulation technique, characterized by a low peak-to-average power ratio, can enhance PUSCH coverage. However, the current specifications restrict pi/2-BPSK usage to only the lowest MCS levels, while higher uplink data rates require the use of higher MCS levels. In Release 20, 3GPP extends the use of pi/2-BPSK modulation to higher MCS levels, thus boosting PUSCH coverage and enabling higher uplink data rates. Another coverage enhancement area in Release 20 is to enable multiple physical random access channel (PRACH) transmissions with different transmit beams. This enhancement enables UE to select optimal uplink beams more effectively, thereby improving cell coverage.

### III. 5G USAGE SCENARIO EXPANSION

5G needs to serve diverse usage scenarios. 3GPP Release 20 continues to study and expand 5G capability in the areas of NTNs, ISAC, ambient IoT, and XR and mobile AI, as illustrated in Figure 3.

*A. Non-terrestrial Networks*

NTNs leverage satellites or high-altitude platforms to provide connectivity services. The integration of NTNs with IoT, referred to as IoT NTN, has garnered considerable commercial interest, particularly through narrowband IoT (NB-IoT) [9]. This area of development has progressed steadily within 3GPP, starting with specifications in Release 17 and continuing with further enhancements in Releases 18 and 19. In Release 20, 3GPP targets enabling voice calls over geosynchronous (GEO) satellites using NB-IoT technology. GEO satellites offer broad global coverage, particularly in regions lacking ground infrastructure, making them ideal for low-data-rate services such as emergency short message services and voice calls. Current solutions for satellite-based communications are largely proprietary. A 3GPP-standardized solution has the potential to enhance interoperability, reduce costs, and address the limitations of existing NB-IoT systems, which were initially designed for IoT use cases and did not support voice services.

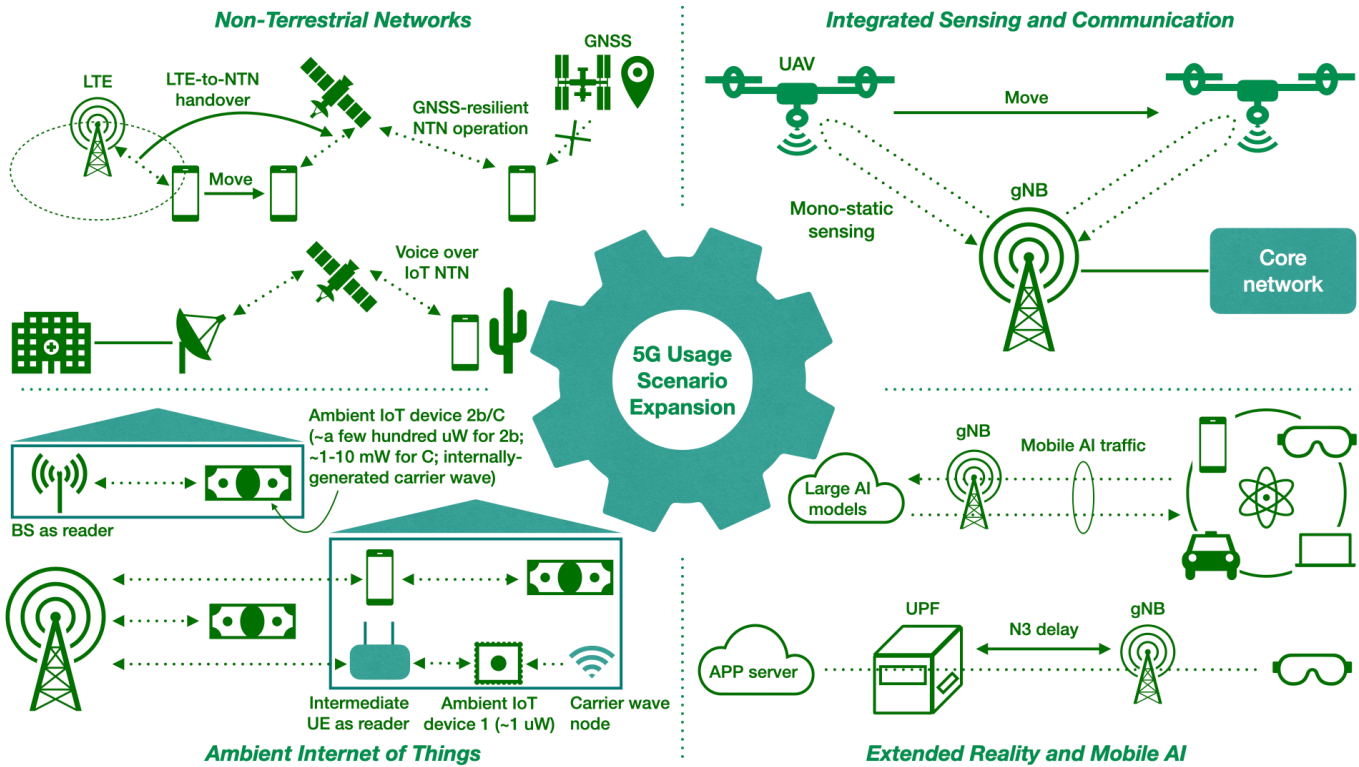

Figure 3: An illustration of the key 3GPP Release-20 features related to 5G usage scenario expansion.

To enable voice calls over GEO satellites using NB-IoT, 3GPP plans to include several key enhancements to NB-IoT in Release 20. Its first task is to down-select between control-plane and user-plane solutions for voice transport. Besides, 3GPP introduces semi-persistent scheduling (SPS) for periodic voice packets in both downlink and uplink communications. SPS can reduce scheduling overhead and delay, making it a preferable approach for voice communication, but SPS is currently not supported in NB-IoT data transmission. The work item also addresses necessary modifications in the RRC connection setup procedure and for emergency call support. Lastly, 3GPP studies the feasibility of increasing UE transmit power, potentially up to 37 dBm, so that uplink voice packets can reach the satellite with a sufficient link margin.

Additionally, 3GPP continues the NR NTN evolution in Release 20. In previous releases, UEs operating via NTN rely on global navigation satellite system (GNSS) positioning data to adjust for time and frequency offsets. However, dependence on GNSS poses issues such as vulnerability to GNSS disruptions, calling for resilient alternatives in cases where GNSS is unavailable or compromised. To this end, 3GPP studies solutions to enable GNSS-resilient NR NTN operation.

Considering the widespread deployment of LTE in terrestrial networks, 3GPP specified mechanisms for idle-mode mobility between LTE terrestrial networks and NR NTN in Release 19. To extend this capability for UEs in connected mode, Release 20 introduces enhancements to inter-radio access technology (RAT) measurements for handover from LTE terrestrial networks to NR NTN. In particular, the work introduces satellite-specific parameters, such as polarization information and ephemeris data, enabling LTE base stations to configure NR NTN neighbor cell measurements properly.

*B. Integrated Sensing and Communication*

ISAC represents a paradigm shift in mobile network technology. It extends mobile networks' traditional communication functionality by integrating sensing capabilities to obtain spatial information about unconnected objects and their surrounding environments [10]. This enables advanced applications such as object detection, tracking, imaging, and mapping.

ISAC has been a focus of academic and industrial research for years, with 3GPP recently taking initial steps toward its standardization. In Release 19, 3GPP conducted a study item that identified deployment scenarios and developed channel models for sensing. The channel modeling effort for ISAC in Release 19 extended the existing 3GPP channel models to account for sensing-specific aspects, including the characteristics of sensing targets and the surrounding background environment. Key developments include radar cross-section modeling, sensing target mobility, deterministic sensing environments, and spatial consistency. These enhancements laid the groundwork for ISAC performance evaluation and standardization.

ISAC is expected to be a fundamental feature of 6G. However, to meet immediate commercial needs, 3GPP continues standardization work on sensing in Release-20 5G-Advanced, focusing on gNB mono-static sensing for the uncrewed aerial vehicle (UAV) use case. The RAN is responsible for sensing signal transmission, measurement, and reception, while the core network handles sensing data and exposes results to third-party applications. While the sensing functionality relies on existing 5G downlink waveform and reference signals, 3GPP studies procedures and signaling between the RAN and core network to support ISAC and investigates network architecture for gNB mono-static sensing. Additionally, 3GPP conducts sensing performance evaluation for gNB mono-static sensing, leveraging the channel models developed in Release 19.

*C. Ambient Internet of Things*

A shift is underway toward IoT technologies powered by ambient energy sources, including radio waves, light, motion, and heat. These IoT devices are capable of operating without batteries or relying on minimal energy storage. However, the output power of energy harvesters, typically ranging from 1 μW to a few hundred μW, is far below the operational requirements of existing cellular IoT technologies (e.g., NB-IoT, LTE machine type communication (LTE-M), and reduced capability (RedCap)), which exceed 10 mW.

To meet the demands of ambient-powered applications, 3GPP has been developing solutions for ambient IoT, a new ultra-low-complexity and ultra-low-power IoT technology designed for deployment in 3GPP systems [11]. In Release 18, 3GPP conducted a study that defined use cases, deployment scenarios, and device characteristics for ambient IoT systems. In Release 19, 3GPP carried out a subsequent study that investigated radio solutions for these devices, emphasizing ultra-low-power designs and new connectivity models. After the study, 3GPP conducted a work item to standardize ambient IoT technology, with a focus on indoor inventory and command use cases. The work targeted ambient IoT device type 1 with ~1 μW peak power consumption, energy storage capability, and a radio frequency (RF) envelope detection receiver. The transmission of ambient IoT device to reader (D2R) relies on backscattering a carrier wave provided externally. With licensed frequency division duplex spectrum in the frequency range (FR) 1, D2R and carrier wave transmissions utilize the uplink spectrum, while reader-to-device (R2D) transmission uses the downlink spectrum. The D2R backscattering supports on-off keying (OOK) and BPSK modulations, whereas the R2D transmission supports only OOK modulation.

In Release 20, 3GPP expands on the Release-19 work to study the support of ambient IoT device types 2b/C with higher peak powers. In contrast to passive device type 1 whose D2R transmission is backscattered on an external carrier wave, the D2R transmission of active device types 2b/C is generated internally by the device. The study addresses the support of device types 2b/C in outdoor scenarios, where outdoor active ambient IoT devices communicate with the outdoor base station reader. Meanwhile, 3GPP introduces specification support for device types 2b/C in indoor scenarios under two topologies: 1) indoor ambient IoT base station as a reader communicates with devices of types 2b/C, and 2) base station communicates with an intermediate UE, which serves as a reader to communicate with devices of types 2b/C. Additionally, the work item specifies the support of device type 1 in indoor scenarios under the second topology; however, the external node providing the carrier wave for the passive device's backscattering is a different node from the intermediate UE reader, as illustrated in Figure 3.

*D. Extended Reality and Mobile Artificial Intelligence*

Previous enhancements for XR in 3GPP Releases 18 and 19 have improved the network's capability to manage XR traffic. One key enhancement area was XR application awareness in 5G networks. To this end, 3GPP introduced the concept of a protocol data unit (PDU) set, which, for example, can be a video frame. The PDU set framework allows the network to handle all PDUs within a set in an integrated manner. In Release 20, 3GPP defines mechanisms to control delay measurement on the N3 interface between the gNB and the user plane function (UPF) in the core network. By accurately measuring N3 delays, the gNB can adjust scheduling strategies and resource allocation to provide better latency guarantees for XR applications.

Additionally, new challenges arise due to the proliferation of mobile AI applications, including large language models (LLMs) and AI-driven image processing. Mobile networks need to evolve to support the rapid growth of mobile AI traffic. In Release 20, 3GPP investigates the transmission characteristics of mobile AI traffic. If justified, 3GPP will specify potential enhancements to better manage mobile AI's uplink traffic, e.g., mobile AI data awareness and the utilization of the PDU set for mobile AI data.

## IV. AI-ASSISTED 5G

AI/ML has emerged as a powerful technology for optimizing complex 5G networks, addressing domains across core network, RAN, UE, and operations and management [12]. Building on the foundational work completed in Releases 18 and 19, 3GPP continues to incorporate AI/ML in Release 20 to enhance the intelligence of 5G-Advanced, as illustrated in Figure 4.

*A. AI/ML for NR Air Interface*

The integration of AI/ML into the NR air interface seeks to enhance the capabilities of 5G-Advanced while laying the groundwork for AI/ML to become a core aspect of 6G. In Release 18, 3GPP initiated a study on AI/ML applications for the NR air interface, focusing on the lifecycle management (LCM) framework and use cases including CSI feedback, beam management, and positioning. The study demonstrated the potential performance gains of AI/ML-based solutions, but also revealed significant challenges, particularly in inter-vendor interoperability and the generalizability of AI/ML models. In Release 19, 3GPP expanded its efforts by introducing normative specifications for AI/ML-based beam management, positioning, and CSI prediction. Additionally, it further studied two-sided AI/ML models for CSI compression, the need for standardized data collection, interoperability, and testability.

In Release 20, 3GPP focuses on specification support for CSI compression with two-sided AI/ML models, addressing inter-vendor training collaboration challenges and developing interoperable solutions. The work aims to specify necessary signaling and mechanisms for spatial/frequency CSI compression without temporal dimensions. This includes a model-pairing handshake, data collection for training, and inference aspects (such as target CSI type, measurement and report configuration, CSI processing criteria and timeline, and CSI report priority rules). Inter-vendor collaboration on two-sided model training considers multiple directions, including 1) fully specified reference model, 2) standardized encoder model structure with parameter exchange, and 3) standardized dataset format/content with dataset exchange. The work also defines requirements for the encoder, along with a specified test decoder, to ensure interoperability. Furthermore, the work addresses the LCM framework for two-sided AI/ML models, including signaling and protocol procedures for functionality and model selection, activation, deactivation, switching, and fallback, as well as the mechanisms required to orchestrate training, inference, and performance monitoring. Additionally, the work addresses standards-based UE data collection, which is important for training and continuous optimization.

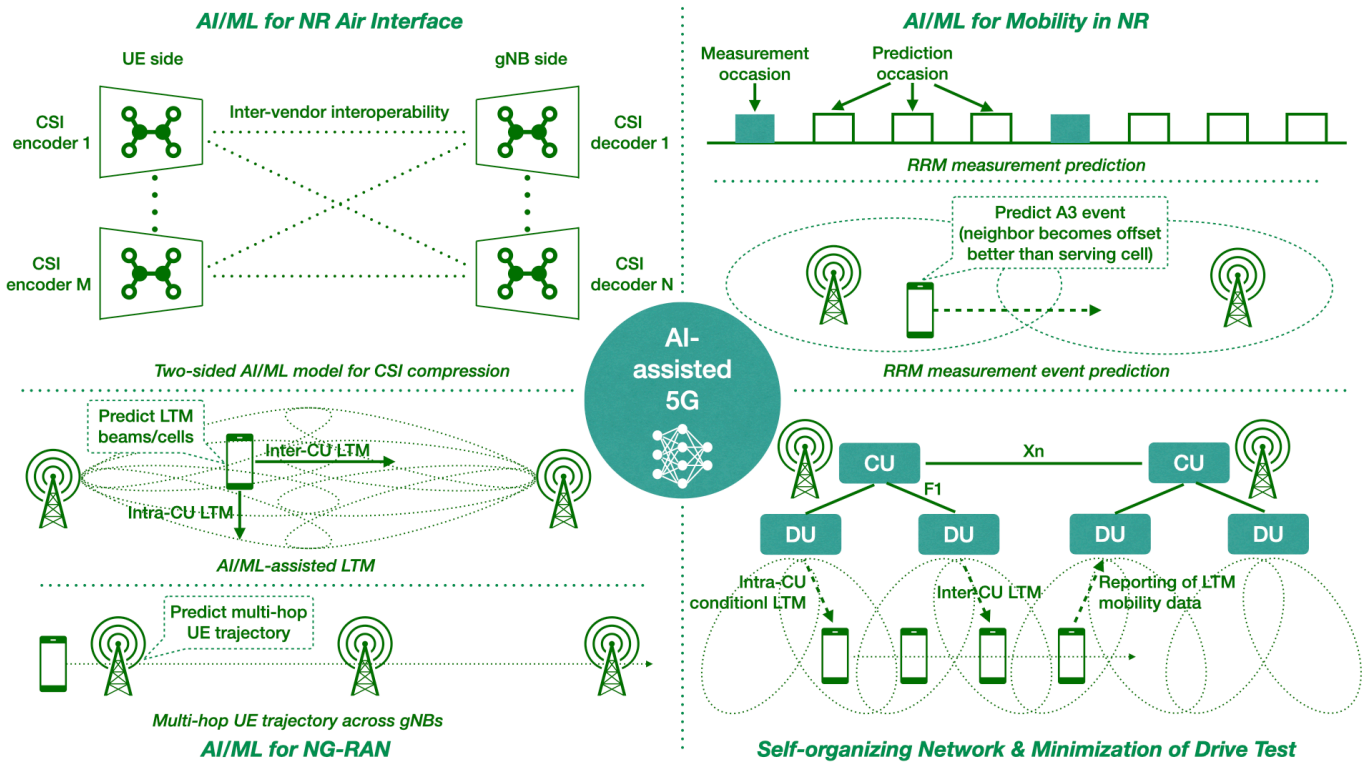

Figure 4: An illustration of the key 3GPP Release-20 features related to AI-assisted 5G.

### B. AI/ML for Mobility

Mobility optimization using AI/ML is another focus area. In Release 19, 3GPP carried out a study item that investigated the potential of AI/ML to enhance mobility solutions by incorporating predictive capabilities and proactive management. Specifically, AI/ML models were investigated for key aspects of mobility, including radio resource management (RRM) measurement prediction, measurement event prediction, and predictions for radio link and handover failures. The evaluation results demonstrated promising gains in prediction accuracy when compared to conventional non-AI/ML methods [13].

Following the Release-19 study, 3GPP aims to introduce specification support for AI/ML-aided mobility in Release 20. The work specifies signaling and protocol aspects for RRM measurement prediction applicable to both UE-side and network-side AI/ML models. The work also introduces signaling and protocol aspects for measurement event prediction for UE-side AI/ML models. Additionally, the work item addresses LCM procedures for the AI/ML models used in mobility management.

### C. AI/ML for NG-RAN

At the next-generation RAN (NG-RAN) architecture level, 3GPP specified enhancements for data collection and signaling support for AI/ML-based network energy savings, load balancing, and mobility optimization in Release 18. In the subsequent Release-19 work item, 3GPP introduced AI/ML support for optimizing network slicing resource allocation and for dynamically adapting cell and beam coverage through cell shaping.

In Release 20, 3GPP continues to study AI/ML-based mobility at the NG-RAN level. Currently, mobility trajectory prediction is limited to the immediate next target RAN node. In Release 20, 3GPP considers expanding prediction to multiple subsequent nodes, which can enhance proactive mobility management. The study also investigates the role of AI/ML in enhancing intra-CU LTM and other handover scenarios, such as inter-CU LTM. The predictive and data-centric approaches have the potential to further increase handover reliability, minimize interruption latency, and optimize the allocation of mobility-related radio resources.

### D. Self-organizing Network and Minimization of Drive Test

Self-organizing network (SON) is an automation technology that supports self-configuration, self-optimization, and self-healing in mobile networks. Minimization of drive test (MDT) is a functionality designed to gather network performance data directly from UEs, reducing the reliance on traditional, labor-intensive drive tests. SON/MDT has been progressively enhanced across multiple releases to support new use cases and accommodate advanced features in the 5G evolution. In particular, SON/MDT encompasses mobility robustness optimization (MRO), which has been enhanced to align with the advancements in 5G mobility across releases.

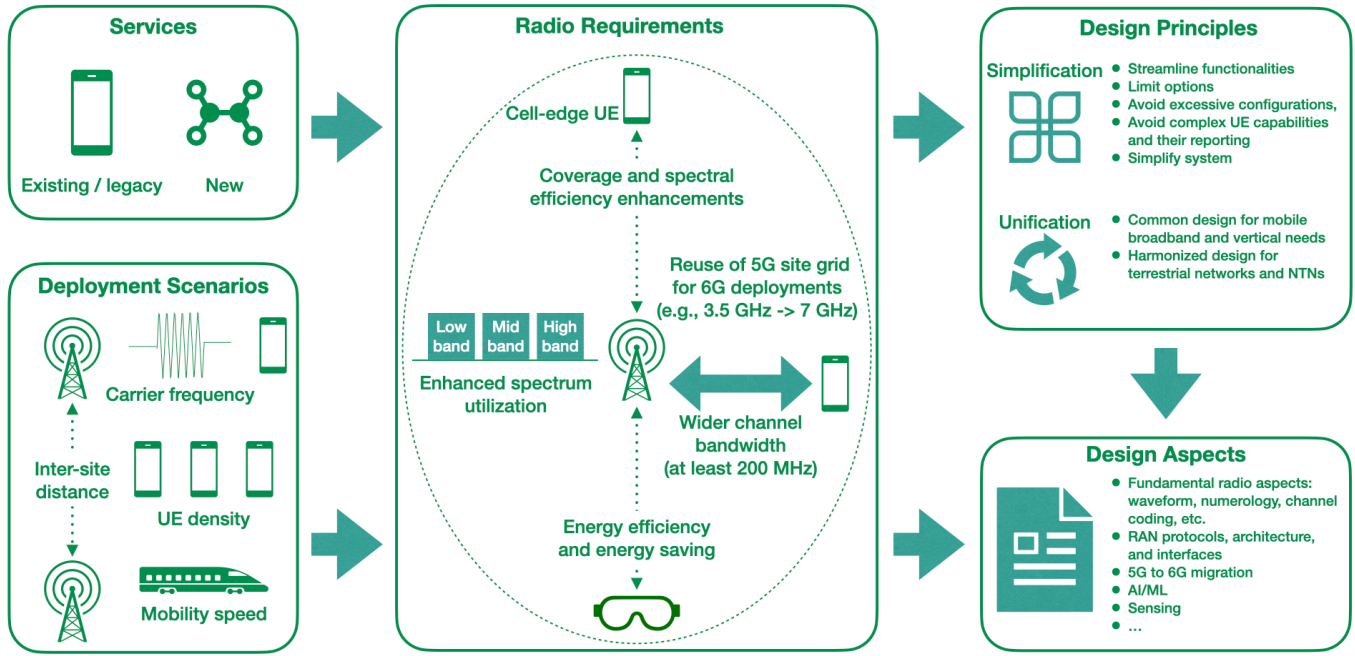

**Figure 5: An illustration of the 3GPP Release-20 studies on 6G RAN.**

MRO solutions typically focus on reporting mobility-related failures experienced by UEs, such as radio link failures. In addition to radio link failure reports, MRO may incorporate historical data tracking the history of cells a UE has visited. In Release 20, 3GPP introduces MRO enhancements for LTM features introduced in Release 19, including inter-CU LTM and intra-CU conditional LTM. To achieve this, the work addresses inter-node information exchange and examines potential enhancements to the existing interfaces. Additionally, the work enhances UE reporting to help the network in tuning mobility parameters.

## V. 6G RADIO ACCESS NETWORKS

The deployment and real-world application of 5G have not only driven digital transformation but also revealed valuable insights regarding network operations, capabilities, complexities, and limitations. Building upon these insights, 6G is expected to further extend network capabilities, opening up new service opportunities while addressing societal needs related to sustainability, reliability, and economic feasibility [14]. In this section, we discuss the 6G RAN study items in 3GPP Release 20, as illustrated in Figure 5.

### A. Scenarios, Requirements, and Principles

In Release 20, 3GPP conducts a 6G RAN study on scenarios and performance requirements [15]. It represents a key step in defining the foundational framework for 6G, aligned with the vision of IMT-2030. One key objective is to define technical performance requirements (TPRs), establish target values, and outline key assumptions to support 6G's requirements, reflecting the anticipated demands of industries and societies in the 2030s.

The study considers practical deployment scenarios. Each scenario incorporates elements such as carrier frequency, inter-site distance, user density, and mobility speed. The study seeks to translate the deployment scenarios into clearly defined radio requirements, addressing both the evolution of existing 5G services and the emergence of novel 6G services. Additionally, the study assesses the applicability of continuing support for legacy services to 6G radio (6GR), while also defining the necessary radio requirements to facilitate seamless service continuity. Specifically, the goal is to achieve substantial improvements across a broad spectrum of network performance indicators, including:

- *Coverage and spectral efficiency enhancements*: Significant improvements in both overall spectral efficiency and cell coverage, placing particular emphasis on cell-edge connectivity and uplink coverage performance.
- *Improved energy efficiency and energy saving*: Prioritizing network-wide energy efficiency improvements for both the network and the device. This aligns with the growing industry priorities around sustainability and operational cost reductions.
- *Deployment evolution*: Ensuring a smooth evolution from existing 5G mid-band (~3.5 GHz) deployments toward new 6G frequency bands (~7 GHz), maintaining comparable coverage and service consistency.
- *Enhanced spectrum utilization*: Targeting optimized spectrum usage, accommodating growth in diverse

- *Wider channel bandwidth*: Supporting wider channels, specifically at least 200 MHz channel bandwidth for 6G deployments in frequencies above 2 GHz and particularly around 7 GHz. The 200 MHz channel bandwidth doubles the maximum channel bandwidth supported by 5G at similar frequencies.

A key principle of 6GR design is to ensure functional efficiency by carefully defining an appropriate and streamlined set of functionalities. This involves minimizing the use of multiple solutions for the same functionality. To reduce complexity and improve manageability, the system should also limit the proliferation of configuration options and prevent excessive requirements on UE capabilities and their reporting. The overall system should aim for design simplification. This includes reducing the complexity of configuration and enabling more efficient management of network elements, such as cells and UEs. Simplification will help lower device and network implementation costs, ease deployment and maintenance efforts, and support scalability and long-term evolution of the network architecture.

Another fundamental principle is to adopt a unified 6GR design that places mobile broadband service requirements at its core, while remaining flexible enough to accommodate a wide variety of use cases from different industry verticals. By ensuring that the same radio design supports both consumer and industry-specific applications, the 6G system will enable efficient deployment and broader adoption across various sectors. Equally important is the harmonization of 6GR design across terrestrial networks and NTNs. The integration of NTNs into 5G came as an afterthought. 6G should support seamless integration from the outset, enabling consistent user experiences and simplifying the implementation of global connectivity solutions.

*B. Design Aspects*

In Release 20, 3GPP further conducts a study on 6GR design aspects, aiming to develop a new, non-backward-compatible RAT with standalone operation to address the scenarios and requirements as defined in [15]. 6GR is scoped to operate across FR1 (up to 7.125 GHz), through the FR1–FR2-1 transition frequency (including ~7 GHz), and into FR2-1 (24.25–52.6 GHz), while frequencies above 52.6 GHz (including terahertz (THz)) are outside the current work scope.

At the physical layer level, the 6GR study uses 5G NR orthogonal frequency-division multiplexing (OFDM) based waveforms and modulation as a baseline for benchmarking against other potential proposals. Similarly, the study assumes 5G NR channel coding schemes (low-density parity check (LDPC) and Polar codes) as a baseline and considers potential extensions to meet 6G requirements. The frame structure needs to be designed to interoperate efficiently with 5G NR, enabling multi-RAT spectrum sharing (MRSS). The study will identify channel bandwidths, avoid defining multiple numerologies (i.e., subcarrier spacings) for the same frequency band, and investigate spectrum utilization and aggregation methods. Physical layer control, data scheduling, hybrid automatic repeat request (HARQ) operation, MIMO, and duplexing schemes will be studied for 6GR. The study will also examine physical layer signals, channels, and procedures, particularly those related to initial access (e.g., synchronization signals, broadcast channel, PRACH, random access procedure, system information delivery, and paging).

The radio-interface protocol architecture of 6GR aims to encompass a high-throughput user plane and a streamlined control plane with optimized RRC states and signaling workflows. The study also investigates access stratum security aspects and how to simplify the radio signaling framework for UE capabilities. Additionally, the protocol design incorporates data transfer to accommodate diverse data types from various use cases, including AI/ML and sensing, ensuring that 6GR can serve as a flexible backbone for future data applications. Mobility management in the 6GR study covers seamless handovers and cell (re-)selection across all RRC states. The 6GR study on core and performance requirements addresses RF, RRM, demodulation and performance, testing, as well as other aspects such as handling irregular channel bandwidths.

At the RAN architecture level, the study encompasses both the RAN-core network functional split and the RAN internal functional split, along with their associated interfaces, protocol stacks, and procedures. Migration and interworking strategies enable gradual evolution from 5G NR to 6GR. 5G-6G MRSS enables both standards to operate concurrently in the same frequency allocations. It can leverage the native flexibility of the 5G NR physical layer to coordinate resource usage. Besides, mobility procedures between 5G NR and 6GR ensure uninterrupted service continuity during the transition.

The study on integrating AI/ML into 6GR builds upon the foundational 5G AI/ML framework. It involves identifying AI/ML use cases by evaluating the trade-offs between performance gains, complexity, and other relevant factors. Once priority use cases are established, the study will explore an extensible AI/ML enabler framework. This framework includes LCM procedures that enable the network to dynamically adapt AI/ML functions to varying service requirements and radio conditions. Complementing the LCM procedures, the framework study also addresses mechanisms for data collection and management. Crucially, the 6GR and associated RAN design should support graceful operation in the absence of AI/ML capabilities. AI/ML-dependent functions should have fallback procedures, allowing the system to maintain baseline service levels using non-AI/ML methods. This dual-mode approach guarantees robust performance.

Finally, the sensing study in 6GR involves physical layer design, including waveform, reference signals, and measurement feedback, and evaluates sensing performance for selected use cases with proper channel models. Additionally, the study addresses the integration of sensing with communication services and investigates higher-layer procedures and protocol aspects. The study further addresses the RF, coexistence, and testability aspects of sensing. Making sensing a native capability of the air interface will equip 6GR with seamless communication and environmental awareness.

## VI. CONCLUSIONS AND FUTURE OUTLOOK

This article has examined the role of 3GPP Release 20 as both a capstone in the evolution of 5G-Advanced and a stepping stone toward 6G. We highlighted how, following the

| Category | Approved/endorsed item | Study/work item | Responsible groups |
|---|---|---|---|
| High Performing 5G | RP-251856: New WI: NR MIMO Phase 6 | Work item | RAN 1, 2, 4 |
| | RP-251865: New WI: NR mobility enhancements Phase 5 | Work item | RAN 2, 4, 3 |
| | RP-251862: New WI: NR coverage enhancements Phase 3 | Work item | RAN 1, 2, 4 |
| 5G Usage Scenario Expansion | RP-251867: New WI: Non-terrestrial networks (NTN) for Internet of Things (IoT) Phase 4 | Work item | RAN 2, 1, 4 |
| | RP-251863: New SI: Study GNSS resilient NR-NTN operation | Study item* | RAN 1, 4 |
| | RP-251878: New WI: E-UTRA TN to NR NTN handover enhancements | Work item | RAN 2, 4 |
| | RP-251861: New SI: Study on integrated sensing and communication (ISAC) for NR | Study item** | RAN 1, 3 |
| | RP-251884: New SI: Study on enhancements for solutions for Ambient IoT (Internet of Things) in NR outdoor for active devices | Study item*** | RAN 1, 4 |
| | RP-251885: New WI: Solutions for Ambient IoT (Internet of Things) in NR Phase 2 | Work item | RAN 1, 2, 3, 4 |
| | RP-251866: New WI: XR (eXtended Reality) for NR Phase 4 | Work item | RAN 2, 3 |
| AI-assisted 5G | RP-251870: New WI: Artificial intelligence (AI)/machine learning (ML) for NR air interface enhancements | Work item | RAN 1, 2, 3, 4 |
| | RP-251864: New WI: Artificial intelligence (AI)/machine learning (ML) for mobility in NR | Work item | RAN 2, 4 |
| | RP-251868: New SI: Study on artificial intelligence (AI)/machine learning (ML) for NG-RAN Phase 3 | Study item** | RAN 3 |
| | RP-251869: New WI: Data collection for SON (self-organising networks)/MDT (minimization of drive tests) in NR Phase 5 | Work item | RAN 3, 2 |
| 6G | RP-251395: Revised SID: Study on 6G scenarios and requirements | Study item | RAN |
| | RP-251881: New SID: Study on 6G Radio | Study item | RAN 1, 2, 3, 4 |

*Note 1: This study item targets 12-month completion and will then be decided whether to proceed with normative work, extend the study item, or postpone normative work to Release 21.
**Note 2: This study item is expected to be followed by a work item in the Release-19 timeframe.
***Note 3: This study item targets 9-month completion and will then be decided whether to include outdoor scenarios in Release-20 ambient IoT work item.

**Table 1: A summary of the 3GPP RAN Release-20 package approved at the June 2025 RAN plenary meeting.**

foundational progress established from Releases 15 through 19, Release 20 adopts a more selective approach, focusing on impactful, deployment-driven enhancements for 5G-Advanced while laying the groundwork for 6G. Table 1 provides a summary of the 3GPP Release-20 package discussed in this article.

As we look ahead, the focus will gradually shift from nurturing the 5G-Advanced ecosystem to shaping the contours of 6G, informed by the lessons learned, architectural frameworks, and validated use cases delivered through 5G-Advanced. Release 21 will lean more decisively into the 6G agenda, representing the first release where 6G studies and early design principles will begin to crystallize. In particular, 3GPP is expected to develop the first set of 6G specifications in Release 21 that will guide the implementations of 6G. The inaugural 6G specifications are expected to be completed in 2029, paving the way for the first commercial 6G deployments in 2030. As the ecosystem progresses, 6G will expand the scope of what mobile communications can achieve.